\documentclass[11pt]{article}

\usepackage{graphicx}
\usepackage{natbib}
\bibpunct{(}{)}{;}{a}{}{,}

\newcommand{\aj}{\textit{Astron.\ Journ.}}
\newcommand{\aap}{\textit{Astron.\ \& Astroph.}}

\newcommand{\apj}{\textit{Astroph.\ Journ.}}  
\newcommand{\mnras}{\textit{Mon.\ Not.\ Roy.\ Ast.\ Soc.}}

\def\la{\mathrel{\mathchoice {\vcenter{\offinterlineskip\halign{\hfil
          $\displaystyle##$\hfil\cr<\cr\sim\cr}}}
    {\vcenter{\offinterlineskip\halign{\hfil$\textstyle##$\hfil\cr
          <\cr\sim\cr}}}
    {\vcenter{\offinterlineskip\halign{\hfil$\scriptstyle##$\hfil\cr
          <\cr\sim\cr}}}
    {\vcenter{\offinterlineskip\halign{\hfil$\scriptscriptstyle##$\hfil\cr
          <\cr\sim\cr}}}}}

\def\ga{\mathrel{\mathchoice {\vcenter{\offinterlineskip\halign{\hfil
          $\displaystyle##$\hfil\cr>\cr\sim\cr}}}
    {\vcenter{\offinterlineskip\halign{\hfil$\textstyle##$\hfil\cr
          >\cr\sim\cr}}}
    {\vcenter{\offinterlineskip\halign{\hfil$\scriptstyle##$\hfil\cr
          >\cr\sim\cr}}}
    {\vcenter{\offinterlineskip\halign{\hfil$\scriptscriptstyle##$\hfil\cr
          >\cr\sim\cr}}}}}

\def\arcmin{\hbox{$^\prime$}}

\begin{document}

\title{Wide--field spectroscopy of A1689 and A1835 with VIMOS: First results}
\author{\Large Oliver Czoske\\\small IAEF, Universit\"at Bonn}

\maketitle

\abstract{ We are using VIMOS to conduct a wide-field spectroscopic
  survey covering fields of $0.5 \times 0.5\,\mathrm{deg^2}$ around
  the X-ray luminous clusters of galaxies Abell 1689 ($z=0.185$) and
  Abell 1835 ($z=0.25$). Here we describe the observations and first
  results on the redshift distribution of subsamples of cluster 
  galaxies to $R\simeq22$ for which we at present have obtained secure
  redshifts. These subsamples constitute $\sim 40\%$ of
  the total spectroscopic sample and contain 525 and 630 cluster
  members in Abell 1689 and Abell 1835, respectively, placing them
  amongst the largest redshift samples available for any cluster of
  galaxies.}

\section{Introduction}
\label{sec:intro}

Imaging surveys of the infall regions of clusters of galaxies which
use photometric redshift techniques to isolate cluster galaxies have
made strong progress in recent years \citep[e.~g.\ ][]{Kodama2001,
  Gray2003}. While photometric surveys have the advantage of being
complete down to well below $M^*$, only spectroscopic surveys can add
a third dimension, velocity, to the galaxy distribution in and around
clusters. The largest wide--field spectroscopic samples at present
exist for near-by clusters, in particular as obtained by the CAIRNS
project \citep{Rines2003}. An example for a previous wide--field
redshift survey at higher redshift is described by \citet{Czoske2001,
  Czoske2002}, who present a catalogue of redshifts for 300 cluster
members with $V \la 22$ in Cl0024+1654 at $z=0.395$. In that case, it
was only the redshift information extending to large cluster-centric
distances which revealed the complex structure of what appeared in
other observations to be a relaxed rich cluster.

The recent advent of high-multiplex spectrographs on 8--10 meter class
telescopes has made it possible to obtain large numbers of
high-quality spectra of galaxies and around clusters of galaxies in a
short amount of time.  The data described by \citet{Czoske2001} were
obtained over the course of four years. Samples larger by a factor of
$2 \dots 3$ can now be obtained in $\sim10\,\mathrm{hours}$ of
observation time. Here I present the first results from a
spectroscopic survey of the two X-ray luminous clusters Abell 1689 and
Abell 1835 at redshifts $z=0.185$ and $z=0.25$ respectively. We use
the VIsible imaging Multi-Object Spectrograph (VIMOS) on VLT
UT3/Melipal. 

\section{Observations}
\label{sec:observations}

The field of view of VIMOS available for spectroscopy consists of four
quadrants of $\sim 7\times 7\, \mathrm{arcmin^2}$, the separation
between the quadrants is $\sim 2\,\mathrm{arcmin}$. Using the LR-Blue
grism, one can place $\sim100\dots150$ slits per quadrant. The
resulting spectra cover the wavelength range
$3700\dots6700\,$\AA\ with a resolution $R\simeq 200$.

We use as the basis for object selection panoramic multi-colour images
obtained with the CFH12k camera on CFHT \citep{Czoske2002diss}, covering
$40\times 30\,\mathrm{arcmin^2}$ in \textit{BRI} for A1689 and
\textit{VRI} for A1835. The input catalogue has been cleaned of stars
(Bardeau et al. 2004, in preparation). 

We attempted to cover the entire CFH12k field of view by using 10
VIMOS pointings for each cluster. Due to technical problems with VIMOS
only 8 and 9 masks were observed in service mode for A1689 and A1835,
respectively. 4 of these masks did not fully meet the observational
requirements although the spectra are at least partly usable. 

In the observed masks, slits were placed on 3373/3542 objects in A1689
and A1835 respectively. The sample was cut at $R=23$, where objects
fainter than $R=22$ were allowed to appear on several masks to provide
higher signal-to-noise for these faint objects. For each mask three
exposures were obtained for a total integration time of 54 minutes.
The data were reduced using the VIPGI pipeline (Scodeggio et al., in
preparation\footnote{\texttt{http://cosmos.mi.iasf.cnr.it/marcos/vipgi/vipgi.html}}). The system subtracts a
bias frame, automatically localises the 2D spectra on the chip and
determines their curvature.  After wavelength calibration from HeNeAr
standard lamp observations the 1D spectra were optimally extracted.

Redshift determination involves an identification step using cross
correlation with a set of templates and a subsequent interactive check
of the correctness of the redshift identification, and a measurement
step using again a correlation with the allowed redshift range
constrained to a narrow interval around the correct identification.
At the time of writing secure redshifts have been determined for 1312/1427
objects in the two clusters. All further results presented in this
contribution will be based on these subsets which reach $R\sim 22$ and
constitute $\sim40\%$ of the total samples.

\section{Redshift space structure}
\label{sec:z-r}

\subsection{Abell 1689}
\label{ssec:A1689-zr}

Abell 1689 is one of the richest ($R=4$) clusters in the Abell
catalogue and shows a complex structure in its inner parts, including
a large number of strongly gravitationally lensed arcs and
arclets. Its X-ray luminosity and temperature are $L_{\rm X} =
5.2\times10^{44}\,h^{-2}\,\mathrm{erg\,s^{-1}}$ \citep{Ebeling1996} and
$T_{\rm X} = 9.3\,\mathrm{keV}$ \citep{Andersson-Madejski2004},
respectively. Based on 96 redshifts in the cluster centre,
\citet{Girardi1997} divide the redshift distribution into three
distinct subgroups. 

\begin{figure}[htbp]
  \centering
  \resizebox{0.48\textwidth}{!}{\includegraphics{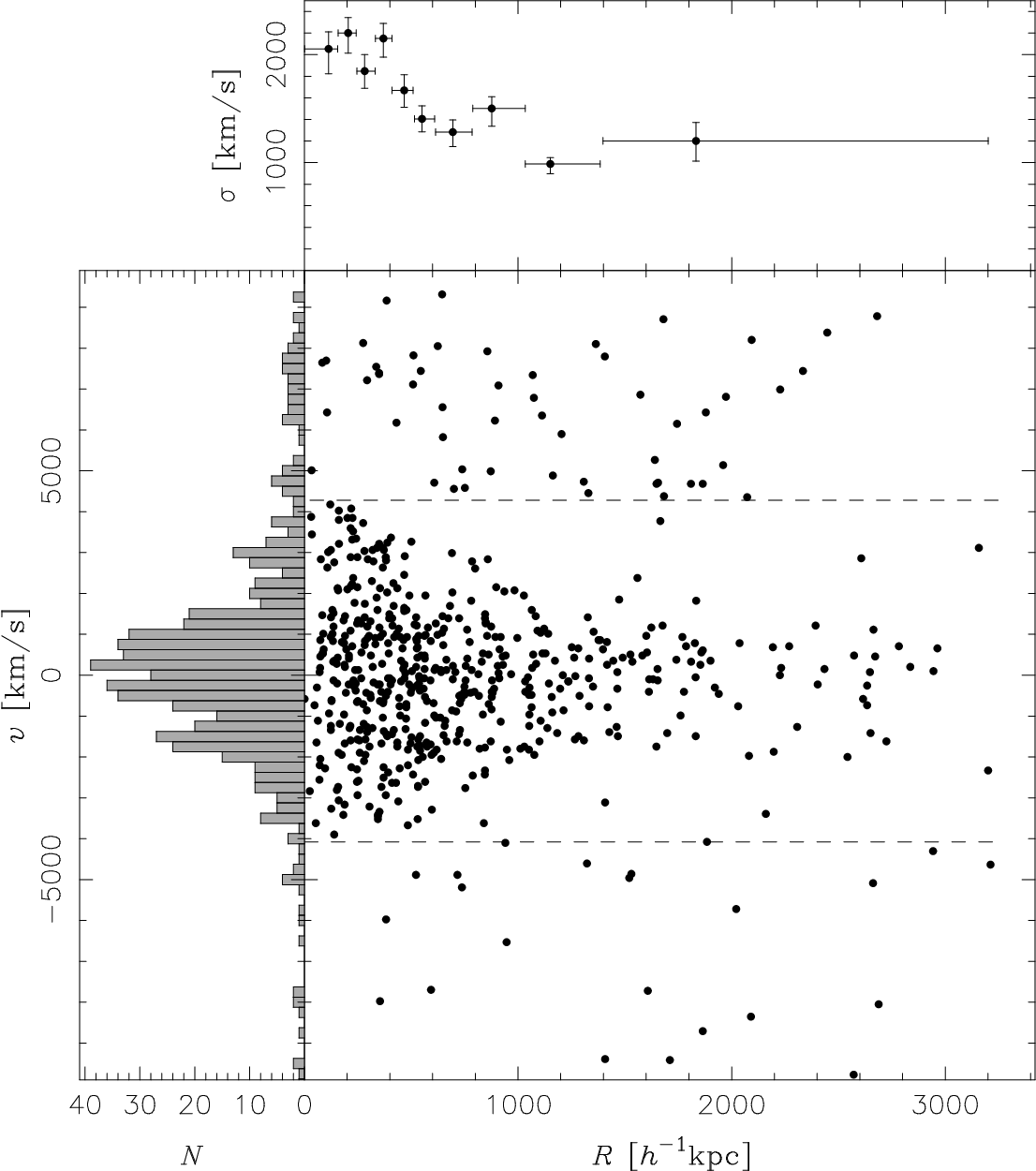}}
  \hfill
  \resizebox{0.48\textwidth}{!}{\includegraphics{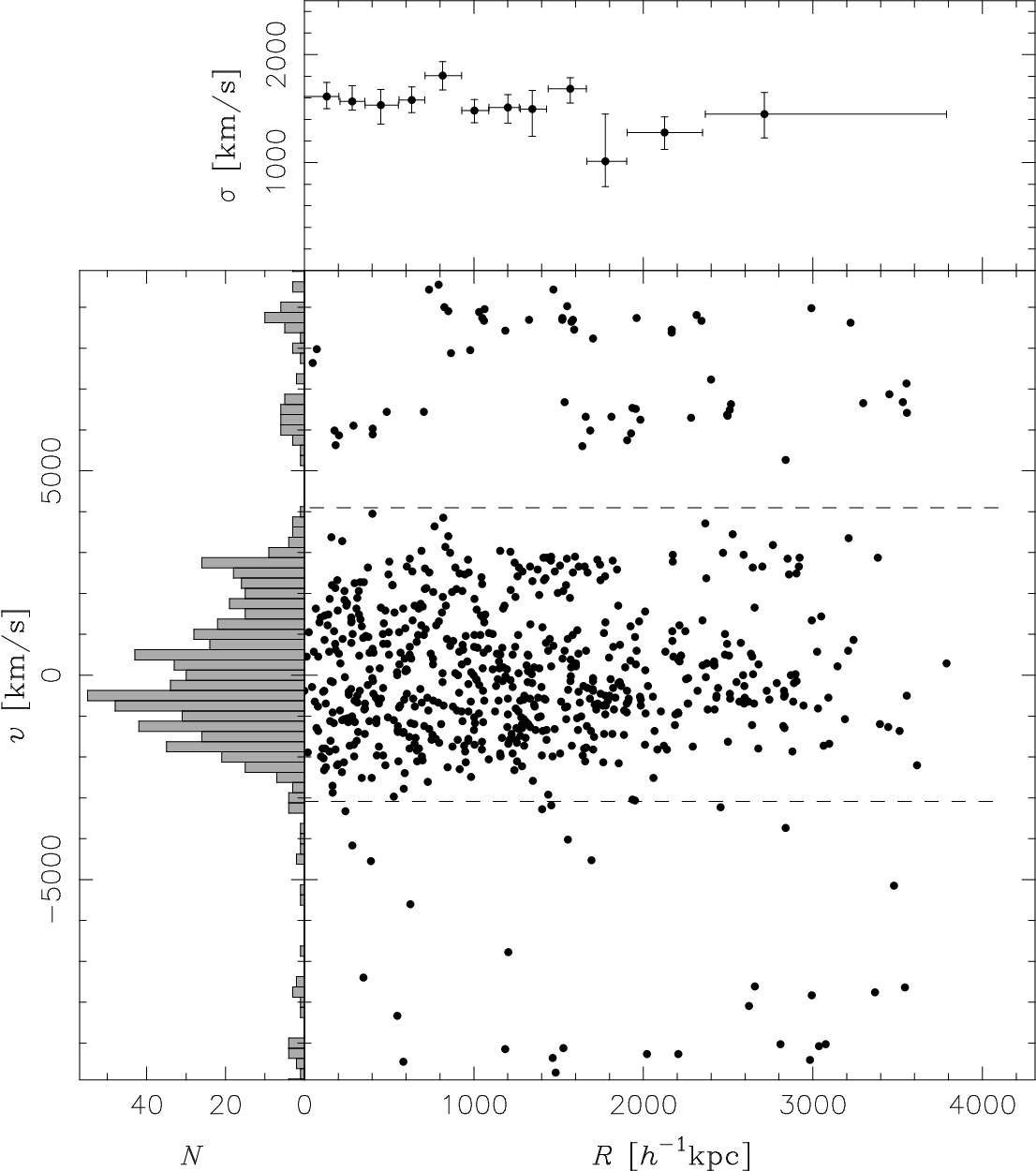}}
  \caption{The redshift distribution for Abell 1689 (left) and Abell 1835 
    (right). Redshifts are given as cosmologically corrected
    velocities relative to the mean cluster redshifts. The velocity
    dispersion profiles in the top panels were computed in bins
    containing 50 galaxies each within the cluster limits marked in
    the central panel.}  
  \label{fig:z-r}
\end{figure}

The left panel of Fig.\ \ref{fig:z-r} shows the redshift distribution
for galaxies near the mean cluster redshift of Abell 1689 as derived
from our VIMOS data. We include redshifts from the literature and
previous observations of the cluster centre using 4m class
telescopes. We define cluster membership by the redshift range $0.169
< z < 0.202$ and find 525 cluster members in the subsample. 

The distribution on large scales is very regular, with clearly defined
``caustics'' whose amplitude decreases strongly with clustercentric
distance. This is also reflected in the velocity dispersion profile
which decreases from $\simeq 2100\,\mathrm{km\,s^{-1}}$ in the central
bins to $\simeq 1200\,\mathrm{km\,s^{-1}}$ for the outer bins. The
redshift histogram in conjunction with the redshift-distance scatter
plot hints at the presence of a distinct subgroup in the cluster
center at a clustercentric velocity of $\sim
3000\,\mathrm{km\,s^{-1}}$ which might artificially enhance the
central velocity dispersion.

\begin{figure}[htbp]
  \centering
  \resizebox{0.45\textwidth}{!}{\includegraphics{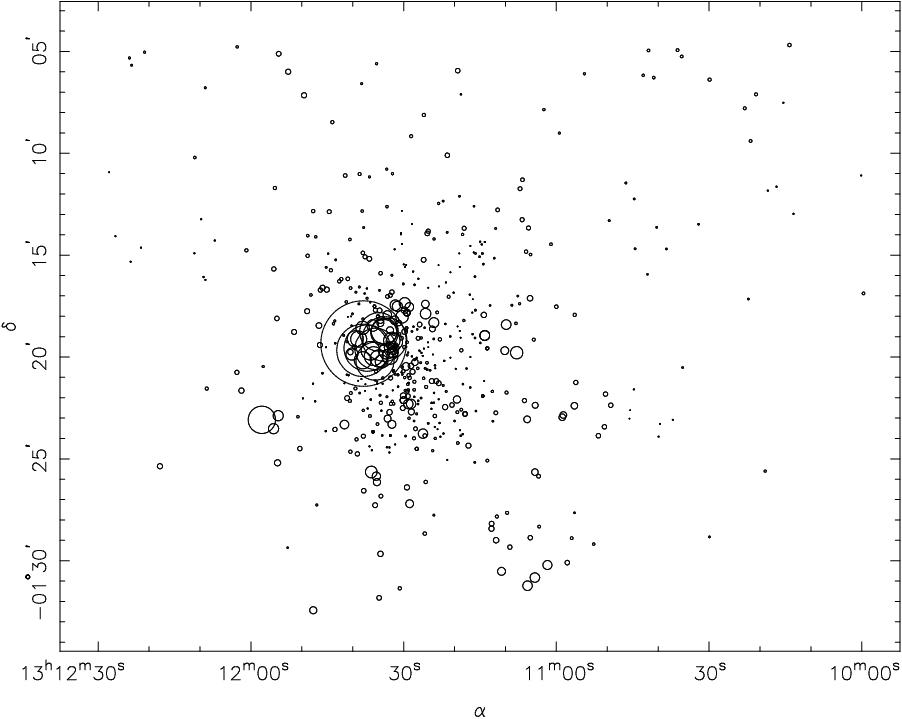}}
  \hfill
  \resizebox{0.45\textwidth}{!}{\includegraphics{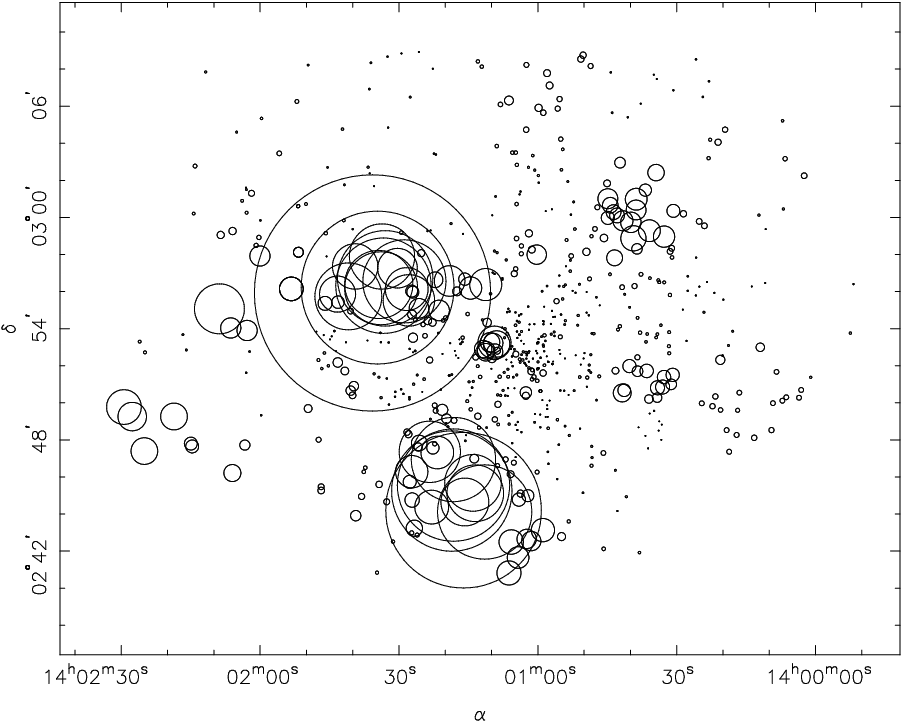}}
  \caption{Dressler-Shectman plots for Abell 1689 (left) and Abell
    1835 (right). The circles have radii proportional to
    $\exp(\delta)$, where $\delta$ is the Dressler-Shectman statistic
    (see text for details).} 
  \label{fig:DS}
\end{figure}

In order to probe for the presence of substructure in Abell 1689 we
apply the test described by \citet{Dressler-Shectman1988}. This test
computes the statistic $\delta \sim \left[\left(\overline{v}_{\rm
      local} - \overline{v}\right)^2 + \left(\sigma_{\rm local} -
    \sigma\right)^2\right]$ for each galaxy to compare the local
velocity distribution based on the 10 nearest neighbours to the global
distribution. The left panel in Fig.\ \ref{fig:DS} shows the result
for Abell 1689. The only apparent distinct group of galaxies lies at
$\sim 2\arcmin$ to the north-east of the cluster center.  The location
of this group, which has been seen in previous redshift samples,
corresponds to a group of bright galaxies in images of the cluster.
The distribution of redshifts of these galaxies is skewed towards
higher redshifts: 51\% of them lie at $z>0.192$, whereas only 13\% of
all cluster members lie at this redshift. The distribution of all
galaxies with $z>0.192$ covers a much larger field but seems
concentrated on the location identified by the Dressler--Shectman
test. \citet{Andersson-Madejski2004} use X-ray data obtained with
XMM/Newton to derive a redshift map for the emitting intra-cluster gas
and find evidence for a higher redshift in their eastern quadrant,
roughly coinciding with our substructure.

\subsection{Abell 1835}
\label{ssec:A1835-zr}

Abell 1835, with $9.6\times10^{44}\,h^{-2}\,\mathrm{erg\,s^{-1}}$, is
the most luminous cluster in the ROSAT Brightest Cluster Sample
\citep{Ebeling1998}. Temperature measurements with ASCA and XMM-Newton
find a comparatively modest global temperature of
$\sim7.5\,\mathrm{keV}$ \citep{Ota2001, Majerowicz2002} with a cool
core of about $4\,\mathrm{keV}$. The surface brightness distribution
is nearly circular to a radius of $\sim0.8\,h{-1}\,\mathrm{Mpc}$. From
observations of the central parts of Abell 1835 it therefore appears
that this is a very massive and relaxed cluster. 

The redshift distribution around Abell 1835 is shown in the right hand
panel of Fig.\ \ref{fig:z-r}. Defining cluster membership by $0.24 < z
< 0.27$, we find 630 cluster members. The distribution of Abell 1835
seems more extended than the distribution of Abell 1689, although a
detailed investigation of the completeness of the samples is necessary
to assess the validity of that statement. The velocity dispersion
profile (VDP) the cluster galaxies in Abell 1835 is flat at $\sim
1500\,\mathrm{km\,s^{-1}}$. From the histogram and the
redshift--distance plot it is clear, however, that the VDP at large
radii includes a contribution from a seemingly separate band of
redshifts at $\sim 3000\,\mathrm{km\,s^{-1}}$ and might thus be
overestimated at $R \ga 1\,h^{-1}\,\mathrm{Mpc}$.

The result of the Dressler-Shectman test on Abell 1835 is shown in the
right hand panel of Fig.\ \ref{fig:DS}. It is immediately apparent
that the infall region is much more structured than the one of Abell
1689, indicating that these clusters inhabit rather different
environments. The Dressler-Shectman test provides evidence for at
least two distinct structures in the outskirts of Abell 1835, one
$\sim 1.5\,h^{-1}\,\mathrm{Mpc}$ to the east of the cluster center,
the other $\sim 1.4\,h^{-1}\,\mathrm{Mpc}$ to the
south-east. The eastern structure is centred on a compact group of
eight galaxies, all with $z>0.26$, significantly higher than the mean
cluster redshift. The southern structure is centred around a bright
elliptical galaxy for which unfortunately no redshift is
available. The mean redshift of the galaxies in this group is 0.245,
somewhat lower than the cluster redshift. 

\section{Conclusions}
\label{sec:conclusions}

The project described in this contribution is still at an early
stage. So far, secure redshifts have been obtained for only $\sim
40\%$ of the observed samples. Nevertheless, the subsamples presented
here include 525 and 630 members galaxies of the clusters Abell 1689
and Abell 1835, respectively. Already, these samples are the largest
available for any cluster with the exception of Coma
\citep{Rines2003}, and demonstrate the power of VIMOS to provide
unprecedented redshift samples for clusters of galaxies at
intermediate redshifts. 

We have described the redshift distribution for the two clusters out
to $\ga 3\,h^{-1}\,\mathrm{Mpc}$ ($\ga 2r_{\rm vir}$). The infall
regions of the two clusters are markedly different, with Abell 1835
showing evidence for strong clustering in its outskirts whereas the
outskirts of Abell 1689 seems much more homogeneous. Curiously, this
is at contrast to the structure of the central parts of the clusters
which is complex for Abell 1689 but seems relaxed for Abell 1835.  The
spatially more concentrated galaxy distribution in Abell 1689 as
compared to Abell 1835 agrees with results from weak lensing analyses
by \citet{Clowe-Schneider2001, Clowe-Schneider2002}, who
find concentration parameters of $c=7.9$ for Abell 1689 and $c=2.9$
for Abell 1835.

The first goal of future work is to complete the redshift
determination for the spectral samples and thus to extend the study of
the cluster structures to fainter magnitude. After proper assessment
of the completeness characteristics of the resulting galaxy catalogues
it will be possible to provide mass profiles based on integration of
the Jeans equation in the inner parts and analysis of the ``caustic''
structure in the outer parts, following \citet{Diaferio1999}, which
can then be compared to mass profiles derived from weak lensing
\citep{Clowe-Schneider2001, Clowe-Schneider2002, Bardeau2004}.  

Further work will relate the spectral types of the galaxies to
cluster-centric distance, the local galaxy density and surface mass
density as determined from weak gravitational lensing. 

\setlength{\bibsep}{0pt}

\end{document}